\documentclass[aps,prl,showpacs,twocolumn,amsmath,amssymb]{revtex4-1}

\usepackage{bm}
\usepackage{graphicx}

\begin{document}

\title{Weak and strong wave turbulence spectra for elastic thin plate}

\author{Naoto Yokoyama}
\email{yokoyama@kuaero.kyoto-u.ac.jp}
\affiliation{Department of Aeronautics and Astronautics, Kyoto University,
 Kyoto 606-8501, Japan}
\author{Masanori Takaoka}
\email{mtakaoka@mail.doshisha.ac.jp}
\affiliation{Department of Mechanical Engineering, Doshisha University,
 Kyotanabe 610-0394, Japan}

\date{\today}

\begin{abstract}
Variety of statistically steady energy spectra 
in elastic wave turbulence have been reported 
in numerical simulations, experiments, and theoretical studies.
Focusing on the energy levels of the system,
we have performed direct numerical simulations according to
the F\"{o}ppl--von K\'{a}rm\'{a}n equation,
and successfully reproduced the variability of the energy spectra 
by changing the magnitude of external force systematically.
When the total energies in wave fields are small,
the energy spectra are close to
a statistically steady solution of the kinetic equation in the weak turbulence theory.
On the other hand,
in large-energy wave fields,
another self-similar spectrum is found.
Coexistence of the weakly nonlinear spectrum in large wavenumbers 
and the strongly nonlinear spectrum in small wavenumbers
are also found in moderate energy wave fields.
\end{abstract}


\maketitle

There exist two types of ``turbulence theories'' started by Kolmogorov and Zakharov.
The former, based on the self-similarity and the dimensional analysis,
is applied on vortical flows
governed by the Navier--Stokes equation.
The latter, based on the random phase approximation,
is applied on wave fields,
and is called the weak turbulence theory (WTT)~\cite{zak_book}.
While Kolmogorov's turbulence theory is applicable to strongly nonlinear turbulence,
WTT,
which is mathematically sophisticated,
is applicable to weakly nonlinear wave fields~\cite{weakturbulence,nls_intermittent_cycle}.

Elastic waves propagating in thin elastic plates
are considered a testing ground
for numerical or experimental verification of the applicability of WTT.
The kinetic equation obtained in WTT
has two statistically steady solutions~\cite{during2006weak}.
One is in a non-equilibrium state,
where the azimuthally-integrated energy spectrum $\mathcal{E}(k)$ as a function of wavenumbers $k$ 
is given as $\mathcal{E}(k) \propto k \left( \log(k_{\ast}/k) \right)^{1/3}$.
The other is in a thermal equilibrium state,
where $\mathcal{E}(k) \propto k$.
The function form of the non-equilibrium solution is determined
so that the energy flux is constant,
but the large-wavenumber cutoff $k_{\ast}$ is indeterminate
in the framework of WTT.
Additionally,
WTT cannot determine the direction of the energy flux 
in the non-equilibrium state in the elastic wave turbulence~\cite{zak_book}.
In these respects,
the elastic wave turbulence is atypical as a weak turbulence system.

D{\"{u}}ring {\itshape et al.\/}~[\onlinecite{during2006weak}] obtained
the energy spectrum corresponding to
$\mathcal{E}(k) \propto k \left( \log(k_{\ast}/k) \right)^{1/3}$
by a direct numerical simulation (DNS).
On the other hand,
energy spectra close to $\mathcal{E}(k) \propto k^{-0.2}$
are reported in experiments using thin elastic steel plates~\cite{boudaoud2008observation,mordant2008there}.
The difference in the energy spectra
is explained in terms of the anisotropy of the system in Ref.~[\onlinecite{PhysRevLett.107.034501}],
and the discreteness of the numerical simulation in Ref.~[\onlinecite{PhysRevE.84.066607}].
Moreover,
dimensional analysis based on the self-similarity predicts
energy spectra corresponding to the energy cascade,
$\mathcal{E}(k) \propto k^{-1}$,
and to the wave action cascade,
$\mathcal{E}(k) \propto k^{-1/3}$~\cite{nazarenkobook}.

We optimistically believe that
there should exist a simple unified explanation 
for the variability of the energy spectra 
including the theoretically-predicted spectra.
In this work,
a series of DNS is performed
according to a basic equation.
Non-equilibrium steady wave turbulent states are obtained
by adding external forces to small wavenumbers
and dissipation to large wavenumbers.
The numerical results provide a unified perspective
on the variability of the spectra.

The governing equation for the lateral displacement $\zeta$
and the momentum $p$
in a thin elastic plate
is called the F\"{o}ppl--von K\'{a}rm\'{a}n (FvK) equation~\cite{fvk},
\begin{subequations}
\begin{align}
&
 \partial_t p = -\frac{Eh^2}{12 (1 -\sigma^2)} \Delta^2 \zeta
 + \left\{ \zeta, \chi \right\}
,
\quad
 \partial_t \zeta = \frac{p}{\rho}
,
\\
&
 \Delta^2 \chi = -\frac{E}{2} \left\{\zeta, \zeta \right\}
,
\end{align}%
\label{eq:zetapchi}%
\end{subequations}%
where $\chi$ is the Airy stress potential.
The Laplace operator and the Monge--Amp\`ere operator
are expressed as $\Delta$ and
$\{f,g\} = \partial_{xx} f \partial_{yy} g + \partial_{yy} f \partial_{xx} g -2 \partial_{xy} f \partial_{xy} g$,
respectively.
The Young's modulus $E$, the Poisson ratio $\sigma$, and the density $\rho$
are the physical properties of the plate.
The thickness of the plate is expressed by $h$.
The FvK equation (\ref{eq:zetapchi}) holds even for large displacements
when the gradient of the displacement is smaller than unity.

The linear dispersion relation
between a wavenumber vector $\bm{k}$ and the corresponding frequency $\omega_{\bm{k}}$
is given as
\begin{align}
 \omega_{\bm{k}} &= \sqrt{\frac{Eh^2}{12 (1 -\sigma^2) \rho}} k^2
,
\label{eq:lineardispersion}
\end{align}
where $k=|\bm{k}|$.
The complex amplitude 
$a_{\bm{k}} = (\rho \omega_{\bm{k}} \widetilde{\zeta}_{\bm{k}} + i \widetilde{p}_{\bm{k}})/\sqrt{2 \rho \omega_{\bm{k}}}$
represents an elementary wave, 
where
$\widetilde{p}_{\bm{k}}$ and $\widetilde{\zeta}_{\bm{k}}$ are
the Fourier components of $p$ and $\zeta$, respectively.
The governing equation can be rewritten as
\begin{align}
 \dot{a}_{\bm{k}} &= - i \omega_{\bm{k}} a_{\bm{k}}
 + \mathcal{N}_{\bm{k}}
,
\label{eq:a}
\end{align}
where
$\mathcal{N}_{\bm{k}}$ symbolically expresses the four-wave nonlinear interaction terms.
Note that no three-wave interactions exist in this system.
(See Ref.~\cite{during2006weak} for the nonlinear terms.)
This equation can also be written in Hamiltonian form,
$i \dot{a}_{\bm{k}} =\delta \mathcal{H} / \delta a_{\bm{k}}^{\ast} $,
where $\delta/\delta a_{\bm{k}}^{\ast}$ denotes
the functional derivative with respect to
the complex conjugate of $a_{\bm{k}}$.

WTT can be applied to the elastic wave turbulence,
if the nonlinear interactions among $a_{\bm{k}}$ are weak
and the random phase approximation is valid.
The wave action $n_{\bm{k}}$ is defined
as
$\langle a_{\bm{k}} a_{\bm{k}^{\prime}}^{\ast} \rangle = n_{\bm{k}} \delta_{\bm{k} \bm{k}^{\prime}}$, 
where $\langle \cdots \rangle$ expresses the ensemble average, 
and $\delta_{\bm{k} \bm{k}^{\prime}}$ is Kronecker's delta.
When the system is statistically isotropic,
we obtain the relation 
$\mathcal{E}(k) \propto k \omega_{\bm{k}} n_{\bm{k}}$.
WTT considers 
the energy transfer due to the weakly nonlinear resonant interactions 
among four wavenumbers.
We have confirmed
that this system has huge number of the quartets which exactly satisfy
the four-wave resonant conditions. 
Then, the system is little affected by the numerical discretization
and it never goes into the frozen turbulence~\cite{pushkarev_1999}.

To investigate the statistically steady states,
the following forcing and dissipative mechanisms are added 
to Eq.~(\ref{eq:a}) as usually done in turbulence simulations.
Energy is inputted at small wavenumbers and dissipated at large ones.
In the forced wavenumbers $|\bm{k}| \leq 8 \pi$,
the absolute value of the complex amplitude $|a_{\bm{k}}|$
is kept constant as $C$ by multiplying the factor $C/|a_{\bm{k}}|$ at each time step
to control the energy of the system easily.
In spite of fixing the absolute value,
the phase of each mode evolves according to Eq.~(\ref{eq:a}).
The artificial eighth-order hyper-viscosity
$- \nu (|\bm{k}|/k_{\mathrm{d}})^8 a_{\bm{k}}$ 
is added to the right-hand side of Eq.~(\ref{eq:a}),
where $\nu$ is a dissipation coefficient
and $k_{\mathrm{d}}$ is a dissipation wavenumber.
The periodic boundary condition is adopted to match the assumptions 
in turbulence theory.
The pseudo-spectral method with the aliasing removal by the 4/2 law
is employed to obtain the nonlinear terms.
The fourth-order Runge--Kutta method is employed for the time integration,
and the linear dispersive term and dissipative term are implicitly solved
to improve numerical stability. 
Independently of the initial conditions,
the system is attracted to a statistically steady state
where the external force and the dissipation balance.

The material properties in the experiment with steel
by Boudaoud {\itshape et al.\/}~\cite{boudaoud2008observation}
are adopted in the present numerical simulations. 
Namely,
$\rho=7.8 \times 10^3$kg/m$^3$, $E = 2.0 \times 10^{11}$Pa, $\sigma = 0.30$,
and $h = 5.0 \times 10^{-4}$m,
respectively.
The plate is supposed to have the periodic boundary $1$m$\times 1$m.

\begin{figure}[t]
 \begin{center}
  \includegraphics[scale=1]{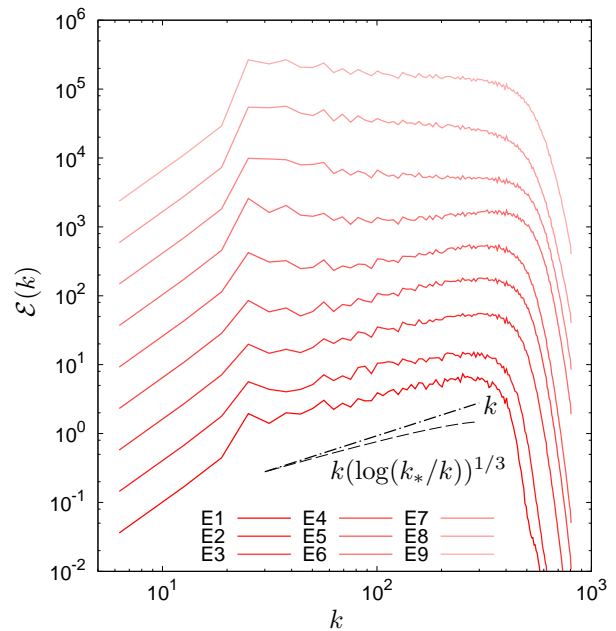}
  \caption{
  Variability of the energy spectra due to energy levels.
  The energy levels are numbered from bottom to top.
  The dashed-dotted line and the broken curve respectively show $k$ and $k (\log(k_{\ast}/k))^{1/3}$
  where $k_{\ast}=144\pi$.
  }
\label{fig:spectrum}
 \end{center}
\end{figure}

A series of DNS with $512^2$ ($256^2$ alias-free) modes~\footnote{
It is not the memory but the CPU time that restricts the number of the modes,
because the time scale of the system is enormously wide
as known from Eq.~(\ref{eq:lineardispersion}).
Moreover,
the time scale of the energy transfer due to the resonance interactions is extremely large in the weak turbulence.
}
is performed in nine energy levels E1--E9.
The energy levels are set
by doubling the constant $C$ in the forcing term.
The ratio of the stretching energy, which is the nonlinear part of the Hamiltonian,
to the bending energy, which is the linear part of the Hamiltonian,
represents the level of the nonlinearity that the entire wave field has.
The ratios are averagely
$1/600$ in E1, $1/20$ in E5, and $1/4$ in E9.
Note that 
the FvK equation is justified
since the root mean squares of the gradients of the displacements $\langle |\nabla \zeta|^2 \rangle^{1/2}$
are respectively $5.0 \times 10^{-3}$ in E1,
$4.3 \times 10^{-2}$ in E5, and $4.8 \times 10^{-1}$ in E9.
The standard deviations of the lateral displacements, $\langle \zeta^2 \rangle^{1/2}$,
are $1.9 \times 10^{-4}$m in E1, $1.4 \times 10^{-3}$m in E5, and $1.4 \times 10^{-2}$m in E9.
The energy spectra for E1--E9 are drawn in Fig.~\ref{fig:spectrum}.
These spectra are time-averaged during the statistically steady states,
and the fluctuations are too small to affect the power-law exponents.
Note that all the spectra in $|\bm{k}| \leq 8\pi$
are proportional to $k^3$ owing to the forcing term.

We observe the spectral similarity
in the low energy levels, i.e., in E1--E3.
In the small energies,
that is, when the nonlinearity is weak, 
the energy spectra approach the spectrum
predicted by WTT, 
$\mathcal{E}(k) \propto k(\log(k_{\ast}/k))^{1/3}$,
which fits the spectra by choosing $k_{\mathrm{d}}=144\pi$ as $k_{\ast}$.
As the energy levels get elevated, i.e., in E4--E6,
the downward-sloping spectrum spreads from the smaller-wavenumber range. 
We observe another spectral similarity in the high energy levels, i.e., in E7--E9,
where the nonlinearity is relatively strong.
The energy spectra are estimated as $\mathcal{E}(k) \propto k^{-0.30 \pm 0.04}$
by the method of least squares.
From now, for simplicity we use the expression ``strongly nonlinear''
for the relatively strongly nonlinear regime where WTT is not justified.
For a wave field with the energy between E6 and E7, 
one may observe $\mathcal{E}(k) \propto k^{-0.2}$
as in Refs.~[\onlinecite{boudaoud2008observation}] and [\onlinecite{mordant2008there}],
if one forcibly fit the spectra to a single self-similar spectrum.
Though their experimental energy spectra
appear to collapse to a similar function form,
the change of the energies are as small as one or two decades.
In the experiment on decaying turbulence
where the change of the energies are relatively large~\cite{PhysRevLett.107.034501},
the energy spectra approach the prediction of WTT
as the energy decays.
Our simulations reproduce all these experimental results
by changing the magnitude of the external force
as shown in Fig.~\ref{fig:spectrum}.
Note that the reproducibility supports the applicability of the FvK equation in all these energy levels
in addition to the spectral similarities mentioned above.

\begin{figure}[t]
 \begin{center}
  \includegraphics[scale=1]{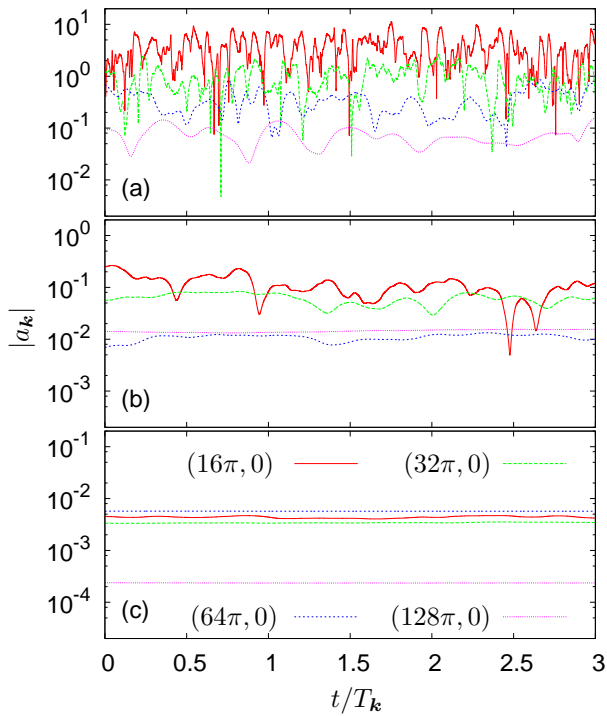}
  \caption{
  The absolute values of the complex amplitudes $|a_{\bm{k}}|$ during three periods.
  (a): Large energy E9,
  (b): moderate energy E5,
  (c): small energy E1.
}
  \label{harmonics} 
 \end{center}
\end{figure}
To investigate the nonlinearities in each scales,
the time evolutions of the magnitudes of elementary waves,
$|a_{\bm{k}}|$,
which would remain constant in the absence of the nonlinear interactions,
are shown in Fig.~\ref{harmonics}.
We select $\bm{k}=(16\pi,0)$, $(32\pi,0)$, $(64\pi,0)$ and $(128\pi,0)$ 
as the representatives in each scales.
Each evolution is drawn for three periods determined by the linear dispersion,
$3T_{\bm{k}} = 6\pi/\omega_{\bm{k}}$.
As the energy becomes larger from E1 to E9,
the fluctuations of $|a_{\bm{k}}|$'s become larger and faster as the overall trend.
In Fig.~\ref{harmonics}(a), i.e., in the large energy E9,
all $|a_{\bm{k}}|$'s are far from constant.
The nonlinear interactions are active in all the scales.
On the other hand,
in Fig.~\ref{harmonics}(c), i.e., in the small energy E1,
all $|a_{\bm{k}}|$'s are almost constant in time.
The nonlinearity at each wavenumber is uniformly weak.
Then,
WTT is applicable for E1, but not for E9.
In Fig.~\ref{harmonics}(b), i.e., in the moderate energy E5,
the smaller-wavenumber waves
have larger and faster fluctuations.
It suggests that
the failure of WTT,
which appears at the wavenumbers
where the linear and nonlinear time scales are comparable~\cite{newell01},
arises in the smaller-wavenumber range.

\begin{figure}[t]
 \begin{center}
  \includegraphics[scale=1]{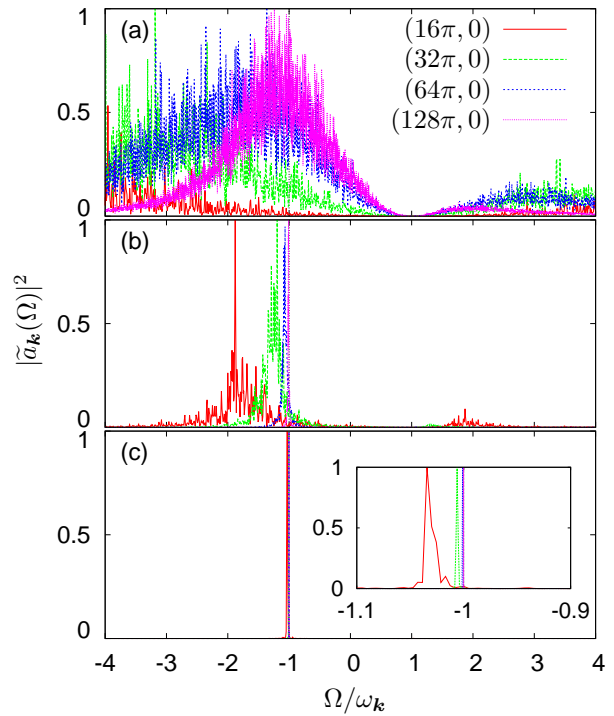}
  \caption{
  The frequency spectrum $|\widetilde{a}_{\bm{k}}(\Omega)|^2$ of each wavenumber.
  (a): Large energy E9,
  (b): moderate energy E5,
  (c): small energy E1, 
  and the spectrum around $\Omega/\omega_{\bm{k}} =-1$ is enlarged in the inset.
  Each spectrum is normalized for visibility so that its maximum is unity.
}
  \label{freqshift} 
 \end{center}
\end{figure}
In Fig.~\ref{freqshift}, 
we examine the frequency spectrum $|\widetilde{a}_{\bm{k}}(\Omega)|^2$
of $a_{\bm{k}}(t)$ at each wavenumber, 
where $\Omega$ is the angular frequency.
The frequency spectra give the conjugate properties 
with the time evolution of $|a_{\bm{k}}(t)|$ shown in Fig.~\ref{harmonics}.
Note that if the nonlinear terms were neglected in Eq.~(\ref{eq:a}),
the frequency spectra would be the line segment at $\Omega/\omega_{\bm{k}} =-1$.
As shown in
Fig.~\ref{freqshift}(a) and (b),
for the higher energy levels and for the smaller wavenumbers,
the frequency spectra become broader,
and
the peaks of the frequency spectrum distribution shift
more to the (negatively) larger frequencies from the linear dispersion relation~\footnote{
The indistinct peak frequency of the mode $(16\pi, 0)$ is found around $\Omega/\omega_{\bm{k}} \approx -6$,
which is out of the range of Fig.~\ref{freqshift}(a).
}.
For the strongly nonlinear waves,
the linear dispersion relation does not characterize the system.
When the energy is small, in Fig.~\ref{freqshift}(c),
the peak frequencies almost coincide with those given by the linear dispersion relation.
Even in this case
the frequency shift to the larger frequencies in the small wavenumbers
are found in the inset.
It is consistent with the observations in the experiments~\cite{PhysRevLett.103.204301},
though the boundary conditions may slightly affect the deviation 
as they described.

The mean frequency shift is estimated by self interactions,
which act as the dispersive term~\cite{nls_intermittent_cycle}.
Then, Eq.~(\ref{eq:a}) can be rewritten as
$\dot{a}_{\bm{k}} = -i (\omega_{\bm{k}} + \Delta \omega_{\bm{k}})  a_{\bm{k}} + \mathcal{N}_{\bm{k}}^{\prime}$,
where $\mathcal{N}_{\bm{k}}^{\prime}$ expresses the non-self nonlinear interactions.
The mean frequency shift due to the self interactions
is given as
\begin{align*}
\Delta \omega_{\bm{k}} = 
\sum_{\bm{k}^{\prime}}
\frac{E}{4\rho^2 \omega_{\bm{k}} \omega_{\bm{k}^{\prime}}}
\frac{|\bm{k} \times \bm{k}^{\prime}|^4}{|\bm{k} - \bm{k}^{\prime}|^4}
\left(|a_{\bm{k}^{\prime}}|^2 + |a_{-\bm{k}^{\prime}}|^2\right)
.
\end{align*}
Suppose that the energy spectrum is self-similar as $\mathcal{E}(k) \propto k^{\alpha}$
and therefore $|a_{\bm{k}}|^2 \propto k^{\alpha-3}$.
Then,
the summation converges if $-1<\alpha<3$,
and the mean frequency shift is proportional to $k^{\alpha-1}$.
Both for the weak turbulence spectrum ($\alpha=1$)
and for the strongly nonlinear spectrum ($\alpha = -0.3$ shown in Fig.~\ref{fig:spectrum})
the mean frequency shift is relatively large in the small wavenumbers
compared with the linear dispersion, $\omega_{\bm{k}} \propto k^2$.
The large mean frequency shift in the small wavenumbers
is consistent with the appearance of the strongly nonlinear spectra in the small wavenumbers
in Fig.~\ref{fig:spectrum}.

We also examine the isotropy of the system
by dividing the wavenumber plane azimuthally into eight regions.
The maximal difference
between the energy in one of the regions and the mean energy in the eight regions
is used as a measure of isotropy.
Even in the highest energy level E9,
the instantaneous relative difference is smaller than $10\%$,
and the time- or ensemble-averaging would make the difference smaller.
Therefore,
the wave field in E9 is statistically isotropic.
The isotropy is validated also by the two-dimensional energy spectrum.
The isotropy in other energy levels is similarly confirmed.
While the deviation of the power-law exponents observed in the earlier studies
from the prediction of WTT
is accounted for by the anisotropy in Ref.~[\onlinecite{PhysRevLett.107.034501}],
the isotropy is maintained in all of our simulations.
We therefore give an alternative mechanism for the deviation,
which is accounted for by the nonlinearity.

In this Letter,
focusing on energy levels,
we have shown that the level of the nonlinearity provides the unified perspective
on the variability of the spectra in the earlier studies~\cite{during2006weak,boudaoud2008observation,mordant2008there,PhysRevLett.107.034501,PhysRevE.84.066607,nazarenkobook}.
The energy spectra in the low energy levels
agree with 
the weak turbulence spectrum $\mathcal{E}(k) \propto k \left( \log(k_{\ast}/k) \right)^{1/3}$. 
Less dissipative plates must be more weakly forced
to reproduce the weak turbulence spectrum in experiments.
The simulations also show another self-similar spectrum 
whose power-law exponent is approximately $-0.30$
in the high energy levels.
The power-law exponent reminds us of $-1/3$
for the ``inverse wave action cascade spectrum~\footnote{Though the
cascade can be confirmed by constant flux, the flux stands little chance
of being obtained in DNS of turbulence.
The direction of the flux might be predicted 
by Fj{\o}rtoft argument,
if the system had two positive conserved quantities to be cascaded.
}'',
which follows from the dimensional analysis,
in Ref.~[\onlinecite{nazarenkobook}].
Note that the physical picture of the wave action is unclear 
in strongly nonlinear wave turbulence
and the wave action is not conserved
even under the kinetic equation in WTT of the present system
owing to the $1\to 3$ and $3\to 1$ asymmetric resonant interactions.
Therefore,
it is still an open question
how the energy spectra in the high energy levels are created.

Moreover,
in the moderate energy levels,
the coexistence of the strongly nonlinear spectrum in the small wavenumbers
and the weakly nonlinear spectrum in the large wavenumbers 
is found.
The coexistence of the weakly and strongly nonlinear turbulence is predicted
in several anisotropic wave turbulence systems~\cite{critical}.
In addition,
the energy equipartition in the small wavenumbers
and the weak turbulence spectrum in the large wavenumbers
are simultaneously observed
in a one-dimensional mathematical model of wave turbulence~\cite{MMT}.
Since the FvK equation describes real physical dynamics,
it is our future work to clarify the relation between 
real-space structures and the fluxes of conserved quantities.

\begin{acknowledgments}
This work was partially supported by the high technological research project
of Doshisha University and MEXT, Japan.
M.T. acknowledges the support in part by KAKENHI Grant No.22540402.
\end{acknowledgments}

\iftrue
\fi

\end{document}